\documentclass[twocolumn,aps,prl] {revtex4} 
\usepackage{graphicx}
\begin{document}
\pagestyle{empty} 
\title{Capillary adhesion between elastic solids with randomly rough surfaces}
\author{  
B.N.J. Persson} 
\affiliation{IFF, FZ-J\"ulich, 52425 J\"ulich, Germany}

\begin{abstract}
I study how the contact area and the work of adhesion, 
between two elastic solids with randomly rough surfaces,
depend on the relative humidity. The surfaces are assumed to be hydrophilic, and
capillary bridges form at the interface between the solids. For elastically hard solids with
relative smooth surfaces, the area of real contact and therefore also the sliding friction,
are maximal when there is just enough liquid to fill out the 
interfacial space between the solids, which typically occurs for $d_{\rm K} \approx 3 h_{\rm rms}$,
where $d_{\rm K}$ is the height of the capillary bridge and $h_{\rm rms}$ the root-mean-square roughness of
the (combined) surface roughness profile.
For elastically soft solids, the area of real contact is maximal for very low humidity 
(i.e., small $d_{\rm K}$), where
the capillary bridges are able to pull the solids into nearly complete contact. In both case,
the work of adhesion is maximal (and equal to $2\gamma {\rm cos}\theta$, where $\gamma$ is the liquid
surface tension and $\theta$ the liquid-solid contact angle) 
when $d_{\rm K} >> h_{\rm rms}$, corresponding to high relative humidity.   
\end{abstract}
\maketitle


{\bf 1. Introduction}

When two solids are in close contact capillary bridges may form at the interface, either as a result
of liquid-like contamination layers (e.g., organic contamination from the normal atmosphere) or water
condensation in a humid atmosphere, or intensionally added thin fluid layers, e.g., thin lubrication
films. For wetting liquids strong negative pressure will prevail in (short) capillary bridges, which will
act as an effective relative long-ranged attraction between 
the solids\cite{IsraelBook}. In many cases, in particular
for hard solids and with rough surfaces, the contribution to the wall-wall 
attraction from capillary bridges may be much larger
than the contribution from the direct wall-wall interaction, e.g., the van der Waals interaction
between the solids. The capillary bridge mediated wall-wall interaction has a huge number of important
applications, e.g., for granular materials\cite{granular}, 
insect or tree frog adhesion\cite{many,frog1,frog2}, head/disk systems\cite{disk} and
microelectromechanical systems (MEMS)\cite{MEMS}, 
where they may trigger permanent adhesion and device failure. 

\begin{figure}
\includegraphics[width=0.47\textwidth,angle=0]{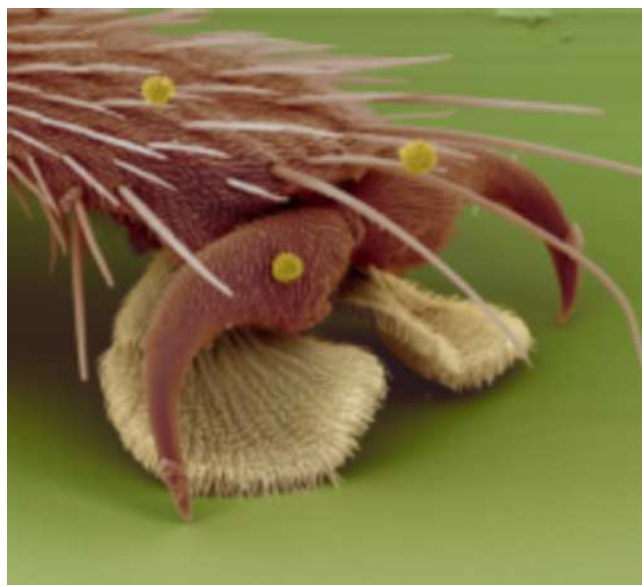}
\caption{\label{HK.fly}
The attachment organ of a fly. At the tip of the hairs on the attachment pads occurs thin
elastic plates. The fly inject a liquid to the space between the plate and the substrate.
The resulting capillary bridge makes it possible for the fly to adhere to surfaces with
large roughness. To very rough surfaces, the fly can also ``adhere'' using the claws on the
legs. Adapted from S. Gorb, with permission.}
\end{figure}

The influence of capillary bridges on adhesion is well known to all of us. Thus, it is possible to build sand
castles from humid or slightly wet sand but not from dry sand or sand flooded with water. Similarly,
it is well known that very flat surfaces, such as those of gauge blocks 
(steel blocks with the roughness amplitude of order $\sim 25 \ {\rm nm}$ when measured
over a macroscopic area, e.g., $50 \ {\rm cm}^2$), adhere with strong forces resulting
from capillary bridges formed from organic contamination or water. Thus it may easily be 
shown that the force to separate two gauge blocks
may strongly increase if one exposes the 
surfaces to (humid) breath (for strong adhesion, the fluid layer (organic or water)
should be at least several times larger than the surface root-mean-square roughness amplitude,
i.e in the present case of order $100 \ {\rm nm}$, or more; see below). 
Finally, the fact that a fly can walk on a vertical
glass window is due to capillary bridges formed at 
the tip of many thin hair-like fibers, which cover the attachment organs of
the fly, see Fig. \ref{HK.fly}. 
Without the fluid (injected via channels in the fibers) adhesion would probably be impossible to most surfaces
because too much elastic energy is necessary to bend the 
tip of the fibers into atomic contact with the rough substrate, which is necessary
for strong adhesion without fluids. 
[The tips of the hair covering the attachment pads of some lizards (e.g., Geckos) and most spiders 
are much thinner that those of a fly, and can easily be bent to make atomic
contact even to very rough surfaces; 
in these cases no liquid is injected by the animal into the contact area.]

Capillary adhesion between solids with randomly rough surfaces
has been studied so far mainly using the Greenwood-Williamson contact
mechanics theory\cite{GW}. 
In this theory the asperities on the rough surfaces are approximated by spherical cups and
the long-range elastic coupling is neglected. It has recently 
been shown\cite{Carbone,BoPersson} that for surfaces with roughness
on many length scales, the GW theory 
(and other asperity contact theories such as the theory of Bush et al\cite{Bush})
fail qualitatively to describe e.g., the area of real contact and the interfacial separation 
as a function of the load. 

When two elastic solids with rough surfaces are squeezed together, the solids will
in general not make contact everywhere in the apparent contact area,
but only at a distribution of asperity contact spots\cite{Borri,Hyun,Chunyan,Carlos}. The separation
$u({\bf x})$ between the surfaces will vary in a nearly random way with the lateral
coordinates ${\bf x}=(x,y)$ in the apparent contact area. 
When the applied squeezing pressure increases, the average 
surface separation $\bar u=\langle u({\bf x})\rangle$ will decrease, but in most situations it is not
possible to squeeze the solids into perfect contact corresponding to $\bar u=0$. 
In thermal equilibrium, capillary bridges are formed at the wall-wall interface in those regions
where the separation $u({\bf x})$ is below the Kelvin length $d_{\rm K}$, which depends on
the relative humidity and on the liquid contact angles with the solid walls.

Here I will present a general theory of capillary adhesion between elastic solids
with randomly rough surfaces. The study is based on a recently developed  
theory for the (average) surface separation $\bar u$ as a function of the squeezing pressure $p$.
The theory shows that for randomly rough surfaces at 
low squeezing pressures $p \sim {\rm exp} (-\alpha \bar u/h_{\rm rms})$
where $\alpha \approx 2$ depends (weakly) on the nature of the surface roughness but is independent of $p$,
in good agreement with experiments\cite{Benz}. The GW contact mechanics theory (and the more accurate theory of
Bush et al) 
instead predicts $p\sim \bar u^{-a}{\rm exp}[-b(\bar u/h_{\rm rms})^2]$ (where $a$ and $b$ are positive numbers)
in strong disagreement with numerical simulations\cite{YangPersson,Pei,PerssonPRL} and experiment\cite{Benz}.

The theory presented below is based on solid and fluid continuum mechanics. Thus it cannot be strictly
applied when the Kelvin distance $d_{\rm K}$ becomes of order the molecular size, i.e., smaller
than $\sim 1 \ {\rm nm}$. The limiting case of molecular thin fluid films must be treated using atomistic
methods, e.g., Molecular Dynamics (see, e.g., Ref. \cite{Samoilov,Yang}).

\begin{figure}
\includegraphics[width=0.40\textwidth,angle=0.0]{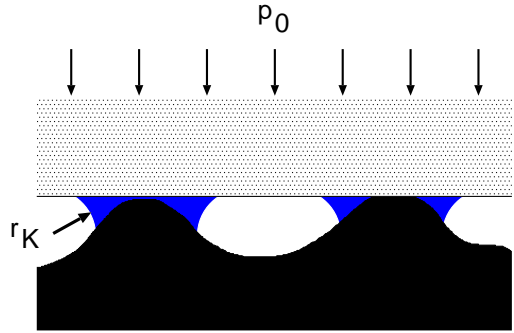}
\caption{\label{capillarypicture}
Capillary bridges formed at two asperity contact areas. At thermal equilibrium, 
the radius of curvature of the
capillary bridge $r_{\rm K}$ is given by the Kelvin radius. The capillary bridges will exert
an attractive force $F_{\rm a}$ on the block. The sum of the capillary force $F_{\rm a}$ and the external load
$F_0 = p_0 A_0$ (where $A_0$ is the nominal surface area) must equal the repulsive force arising
from the area of real contact between the solids.
}
\end{figure}

\begin{figure}
\includegraphics[width=0.45\textwidth,angle=0]{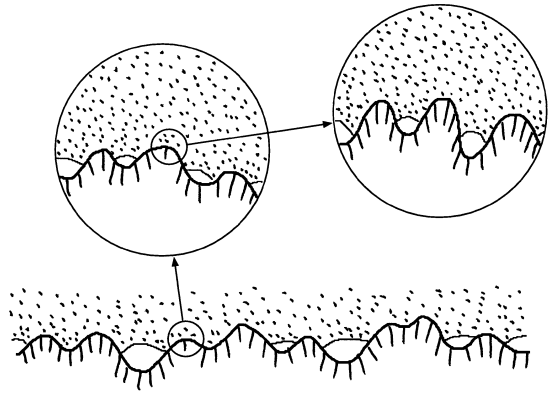}
\caption{\label{1x}
An rubber block (dotted area) in adhesive contact with a hard
rough substrate (dashed area). The substrate has roughness on many different 
length scales and the rubber makes partial contact with the substrate on all length scales. 
When a contact area 
is studied at low magnification it appears as if complete contact occur, 
but when the magnification is increased it is observed that in reality only partial
contact occur.  
}
\end{figure}

\begin{figure}
\includegraphics[width=0.40\textwidth,angle=0.0]{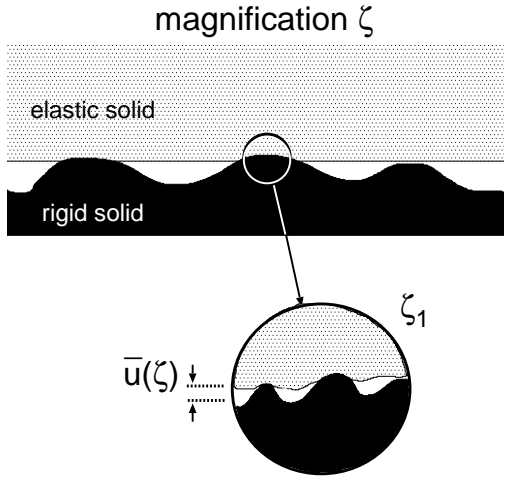}
\caption{\label{asperity.mag}
An asperity contact region observed at the magnification $\zeta$. It appears that
complete contact occur in the asperity contact region, but upon increasing the magnification
to the highest resolution (magnification $\zeta_1$) 
it is observed that the solids are separated by the (average) distance $\bar{u}(\zeta)$.
}
\end{figure}

\vskip 0.5cm
{\bf 2. Theory}

Consider the frictionless 
contact between two elastic solids with the Young's elastic modulus $E_1$ and $E_2$ and the Poisson ratios $\nu_1$ and $\nu_2$.
Assume that the solid surfaces have the height profiles $h_1 ({\bf x})$ and $h_2({\bf x})$, respectively. The elastic
contact mechanics for the solids is equivalent to those of a rigid substrate with the height profile $h({\bf x}) = h_1({\bf x})+
h_2({\bf x})$ and a second elastic solid with a flat surface and with the Young's modulus $E$ and 
the Poisson ratio $\nu$ chosen so
that\cite{Johnson2}
$${1-\nu^2\over E} = {1-\nu_1^2\over E_1}+{1-\nu_2^2\over E_2}.\eqno(1)$$ 

Consider an elastic block with a flat surface in contact with a hard rough substrate.
We consider humid condition (vapor pressure $P_{\rm v}$) and assume that the liquid wet the surfaces.
In this case in the asperity contact regions, liquid capillary bridges will form 
(see Fig. \ref{capillarypicture}), where the 
meniscus radius $r_K$ is given by the Kelvin equation 
$$r_{\rm K}= - {\gamma v_0 \over k_{\rm B} T {\rm ln} (P_{\rm v}/P_{\rm sat})},\eqno(2)$$
where $v_0$ is the molecular volume in the liquid. 
Thus, the thickness of the liquid film is given by
$$d_{\rm K} = r_{\rm K} ({\rm cos}\theta_1+{\rm cos}\theta_2 ),\eqno(3)$$
where $\theta_1$ and $\theta_2$ are the liquid contact angles on the two
solid surfaces. 

For water at $T=300 \ {\rm K}$, $\gamma = 0.073 \ {\rm J/m^2}$ and 
$v_0 \approx 3\times 10^{-29} \ {\rm m}^3$ so that
$$r_{\rm K} \approx - {0.53 \ {\rm nm} \over {\rm ln} (P_{\rm v}/P_{\rm sat})}.$$
Thus, if water wet the surfaces (i.e., $\theta_1=\theta_2=0$) 
the thickness of the liquid film is given by the Kelvin length 
$$d_{\rm K} \approx - {1.06 \ {\rm nm} \over {\rm ln} (P_{\rm v}/P_{\rm sat})}.\eqno(4)$$

We now use the contact mechanics 
formalism developed elsewhere\cite{PSSR,YangPersson,P1,Bucher,JCPpers},
where the system is studied at different magnifications $\zeta$, see Fig. \ref{1x}.
When the system is studied at the magnification $\zeta$ it appears as if the contact area
(projected on the $xy$-plane) equals $A(\zeta)$, but when the magnification
increases it is observed that the contact is incomplete, and the surfaces in the apparent
contact area $A(\zeta)$ are in fact separated by
the average distance $\bar u(\zeta)$, see Fig. \ref{asperity.mag}.
Let $u_1(\zeta)$ be the (average) height separating the surfaces which appear to come into 
contact when the magnification decreases from $\zeta$ to $\zeta-\Delta \zeta$, where $\Delta \zeta$
is a small (infinitesimal) change in the magnification. $u_1(\zeta)$ is a monotonically decreasing
function of $\zeta$, and can be calculated from $\bar u(\zeta)$ and $A(\zeta)$ using
(see Ref. \cite{YangPersson})
$$u_1(\zeta)=\bar u(\zeta)+\bar u'(\zeta) A(\zeta)/A'(\zeta).\eqno(5)$$
We assume that liquid occurs in the apparent contact areas when the
separation $u_1(\zeta)$ is smaller than (or equal to)
the Kelvin distance $d_{\rm K}$. Assume that 
this occurs for the magnification $\zeta=\zeta_{\rm K}$ so that
$$u_1(\zeta_{\rm K},p_0) = d_{\rm K},\eqno(6)$$ 
where we have indicated that the separation $u_1$ depends on the nominal pressure $p_0$.
In the liquid bridge is a negative pressure $p=-p_{\rm K}$
with $p_{\rm K}=2 \gamma /d_{\rm K}$. The liquid occupy the (projected) surface area 
$\Delta A = A(\zeta_{\rm K})-A(\zeta_1)$, 
where $A(\zeta_1)$ is the area of real contact observed at the highest magnification $\zeta_1$.
Thus the attractive force 
$$F_{\rm a} = p_{\rm K} \Delta A = {2\gamma \over d_{\rm K}} 
\left [A(\zeta_{\rm K})-A(\zeta_1)\right ].\eqno(7)$$
We define $p_{\rm a}=F_{\rm a}/A_0$. 
We will calculate $A(\zeta)$ by using a mean-field type of approximation,
where instead of including the non-uniform distribution of capillary forces acting at the block-substrate
interface, we assume that the effective 
squeezing pressure is $p=p_0+p_{\rm a}$, where $p_0=F_0/A_0$ is the applied squeezing pressure
(which is negative during pull-off) (see Fig. \ref{capillarypicture} and also Sec. 5). 

Let us apply a pull-off force $F_0=-F_{\rm pull}$ to the block.
We define the work of adhesion per unit area as
$$w={1\over A_0}\int_{u_0}^\infty d u \ F_{\rm pull}(u)
=\int_{u_0}^\infty d u \ [p_{\rm a}-p(u)].\eqno(8)$$
$p(u)$ is the repulsive pressure from the substrate at the separation $u=\bar u$ 
between the average substrate surface plane and the average position
of the bottom surface of the block. 
$u_0$ is the equilibrium separation when $F_{\rm pull}=0$, i.e.,
$p(u_0)=p_{\rm a}$.

Let us now study the limiting case when the space between the solids is filled with liquid, i.e.,
no dry area. We also assume that the area of real contact $A(\zeta_1)$ is negligible compared to
the nominal contact area $A_0$. In this case the attractive pressure $p_{\rm a} = 2\gamma /d_{\rm K}$ 
must be balanced by the repulsive asperity contact pressure which for 
separation $u \gtrsim h_{\rm rms}$ is given by
$$p(u)= \beta E^* e^{-\alpha \bar u / h_{\rm rms}}.\eqno(9)$$
where $\alpha$ and $\beta$ are numbers which depend on the surface roughness
but which are independent of $p$ and of the elastic properties of the solids\cite{PerssonPRL}. Thus
$${2\gamma \over d_{\rm K}} = \beta E^* e^{-\alpha u_0 / h_{\rm rms}},\eqno(10)$$
or 
$$u_0= {h_{\rm rms} \over \alpha} {\rm log} {\beta E^* d_{\rm K} \over 2 \gamma}.\eqno(11)$$
Since the capillary pressure is 
$p_{\rm a} = 2 \gamma /d_{\rm K}$ for $u<d_{\rm K}$ and zero otherwise, we get work of adhesion
$$w=\int_{u_0}^\infty d u \ [p_{\rm a}-p(u)]$$
$$ = {2\gamma \over d_{\rm K}} \left (d_{\rm K}-u_0\right )
-{\beta E^* h_{\rm rms}\over \alpha}e^{-\alpha u_0 / h_{\rm rms}}.\eqno(12)$$
Using (11) this gives
$$w=2\gamma \left [1-{h_{\rm rms} \over \alpha d_{\rm K}}{\rm log} \left ({e\beta E^* d_{\rm K}\over 2\gamma}
\right )\right ].\eqno(13)$$
For self-affine fractal surfaces we have\cite{PerssonPRL} $\alpha \approx 2$ 
and $\beta \approx 0.4 q_0 h_{\rm rms}$
giving
$$w \approx 2\gamma \left [1-{h_{\rm rms} \over 2 d_{\rm K}}{\rm log} \left ({q_0h_{\rm rms} E^* 
d_{\rm K}\over 2\gamma} \right )\right ],\eqno(14)$$
where $q_0$ is the roll-off (or cut-off) 
wavevector of the surface roughness power spectrum.

The analysis above has assumed that thermal equilibrium occurs at any 
moment in time during pull-off, which requires a static or slowly propagating debonding crack.
In this case liquid will condense or evaporate at the interface in such a way as to always maintain
the Kelvin radius $r_{\rm K} = d_{\rm K}/2$ for the radius of curvature of the liquid meniscus.
However, during fast pull-off negligible condensation or evaporation occurs, 
and the fluid volume at the interface,
rather than the Kelvin radius, will be constant during pull-off. 
If $A(u)$ is the area covered by liquid when the average surface separation is $u$, then volume conservation
requires $A_0u_0 = A(u)u$ or $A(u)/A_0 = u_0/u$.
In this case
$$\int_{u_0}^\infty du \ p_{\rm a}(u) =  \int_{u_0}^\infty du \ {2 \gamma \over u}
{A(u)\over A_0} =2\gamma. $$ 
When thermal equilibrium occurs during pull-off the same integral becomes $2\gamma (1-u_0/d_{\rm K})$. 
Thus, when $u_0/d_{\rm K} << 1$, the whole interface is filled with fluid, and the adiabatic pull-off (constant meniscus radius)
and fast pull-off (constant fluid volume) give nearly the same result 
for the work of adhesion, assuming that one may neglect
viscous energy dissipation in the fluid during the fast pull-off (which is responsible for suction-cup type of effective adhesion).

\begin{figure}
\includegraphics[width=0.47\textwidth,angle=0]{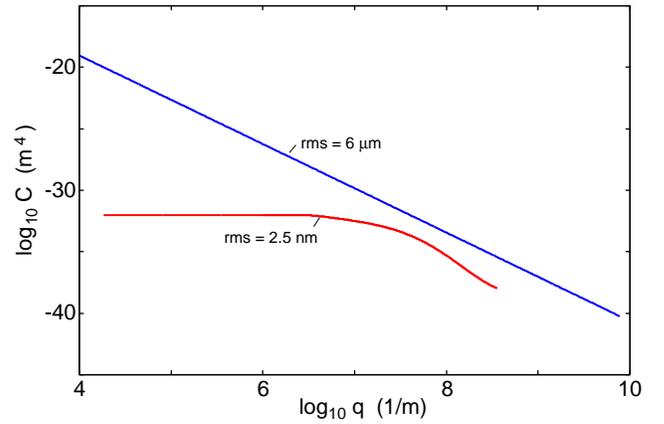}
\caption{\label{Cq.both}
The surface roughness power spectrum of two surfaces. The surface with the root-mean-square (rms)
roughness $6 \ {\rm \mu m}$ is self affine fractal with the fractal dimension $D_{\rm f} = 2.2$.
The other surface has the rms roughness $2.4 \ {\rm nm}$ and is assumed to be randomly rough.
}
\end{figure}

In general, for most solids with covalent, ionic or metallic bonds
$E\approx 10^{11} \ {\rm Pa}$, and in the normal atmosphere
(where $d_{\rm K}$ is of nanometer size) capillary adhesion will only be observed if $h_{\rm rms}$ is
at most a few nanometer. In fact, assuming that (14) is valid we get $w=0$ if 
$$d_{\rm K } < d_c =  {h_{\rm rms} \over 2} { \rm log} \left ({q_0h_{\rm rms} E^* 
d_c \over 2\gamma} \right )\eqno(15)$$
Since $d_c$ only depends logarithmically on the parameters $q_0$, $E^*$ and $\gamma$, 
the critical Kelvin distance (or humidity), 
below which the adhesion is very small, depends weakly on
these parameters.
For example, 
if $h_{\rm rms} \approx 2 \ {\rm nm}$, 
and with the roll-of wavevector $q_0 \sim 10^7 \ {\rm m}^{-1}$ (as for the lower curve in Fig. \ref{Cq.both})
we get ${\rm log}(q_0 h_{\rm rms} E^* d_{\rm K}/2\gamma ) \approx 5$ so 
that $d_{\rm c} \approx 5 \ {\rm nm}$ corresponding to $\sim 80\%$ relative humidity.

As another application of (14), let us study the case where
$E^*=1 \ {\rm GPa}$ (as typical for glassy polymers) in contact with a hard rough substrate
with $h_{\rm rms} = 1 \ {\rm \mu m}$ and $q_0 = 10^4  \ {\rm m}^{-1}$. 
Using $\gamma \approx 0.1 \ {\rm J/m^2}$ (14) gives that $w>0$ only if $d_{\rm K} > 10 \ {\rm \mu m}$.
More generally, (14) shows that capillary adhesion will only manifest itself as long as the
the Kelvin length $d_{\rm K}$ is (at least) a few times the root-mean-square amplitude
of the (combined) substrate roughness profile. This statement only holds for elastically hard enough
solids--for very soft solids the negative capillary pressure can pull the solids into close contact
in which case (14) is no longer valid, see Sec. {\bf 3.2}.

\vskip 0.5cm
{\bf 3. Numerical results}

We will apply the theory to two cases, namely the contact between elastically hard solids,
e.g., silicon as relevant for microelectromechanical systems (MEMSs), and elastically soft systems such
as rubber or biological adhesive pads.  
For the case of hard solids we will assume that the (combined) rough surface has the 
surface roughness power spectrum $C(q)$ shown in Fig. \ref{Cq.both}, bottom curve. This power spectrum was
obtained from the surface topography, $h({\bf x})$,  measured\cite{Private} 
for a polysilicon surface, using\cite{P3}:  
$$C(q) = {1\over (2\pi )^2} \int d^2x \ \langle h({\bf x}) h({\bf 0})\rangle e^{-i{\bf q}\cdot {\bf x}}.$$  
The root-mean-square (rms) roughness of this surface
is $2.4 \ {\rm nm}$. For the case of elastically soft solids we will assume that the combined surface
has the power spectrum shown in Fig. \ref{Cq.both}, top curve. This is the power spectrum of a
self affine fractal surface with the rms roughness $6 \ {\rm \mu m}$ and the fractal dimension $D_{\rm f} = 2.2$.

\begin{figure}
\includegraphics[width=0.47\textwidth,angle=0]{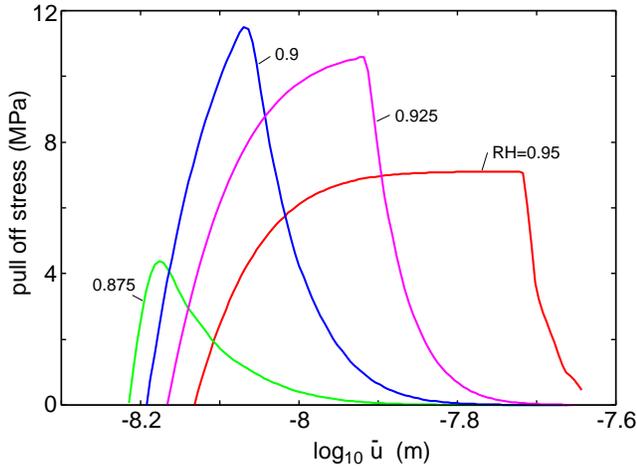}
\caption{\label{pull.off.de.Boer}
The stress as a function of the logarithm of the (average) separation $\bar u$ during 
separation. The area under the curves determines the work of adhesion. For
a hard surface with the root mean square roughness $h_{\rm rms} = 2.4 \ {\rm nm}$
in contact with an elastic solid (with the Young's modulus $E=82 \ {\rm GPa}$ and Poisson ratio
$\nu = 0.22$) with a flat surface. Results are shown for the relative humidities (RH) 0.95, 0.925,
0.9, 0.875. 
}
\end{figure}

\begin{figure}
\includegraphics[width=0.47\textwidth,angle=0]{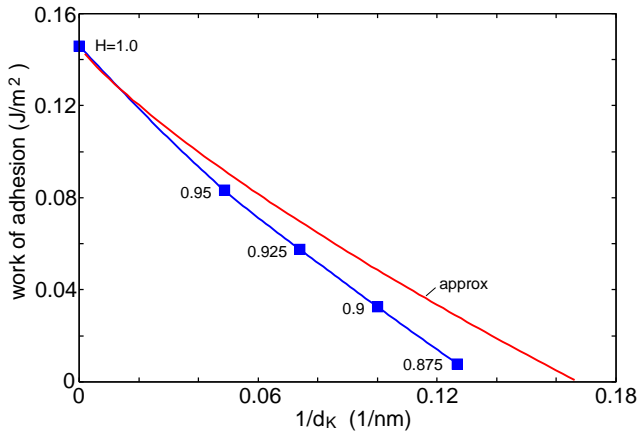}
\caption{\label{DeBoer.w.invers.dK}
The work of adhesion as a function of the inverse of the height $d_{\rm K}$ of the capillary bridges. For
a hard surface with the root mean square roughness $h_{\rm rms} = 2.4 \ {\rm nm}$
in contact with an elastic solid (with the Young's modulus $E=82 \ {\rm GPa}$ and Poisson ratio
$\nu = 0.22$) with a flat surface. The square data points are the calculated results 
for the relative humidities (RH) 0.95, 0.925,
0.9, 0.875, and the solid line a fit to the data. 
The line denoted by ``approx'' is given by (13) with $\alpha = 1.87$ as calculated\cite{YangPersson} from
the measured surface roughness power spectrum.
}
\end{figure}

\begin{figure}
\includegraphics[width=0.47\textwidth,angle=0]{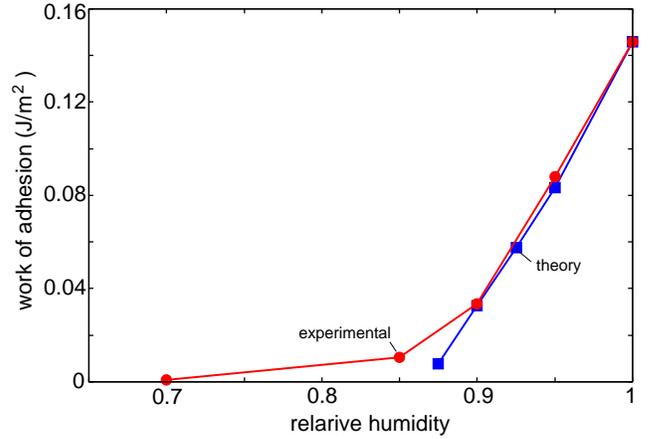}
\caption{\label{theory.exp.RH.w}
The work of adhesion as a function of the relative humidity. The circle are experimental data 
(from Ref. \cite{Frank2}) and the squares
calculated results. In the calculation we have assumed
a hard surface with the root mean square roughness $h_{\rm rms} = 2.4 \ {\rm nm}$
in contact with an elastic solid (with the Young's modulus $E=82 \ {\rm GPa}$ and Poisson ratio
$\nu = 0.22$) with a flat surface. The square data points are the calculated results 
for the relative humidities (RH) 0.95, 0.925,
0.9, 0.875.  
}
\end{figure}

\begin{figure}
\includegraphics[width=0.47\textwidth,angle=0]{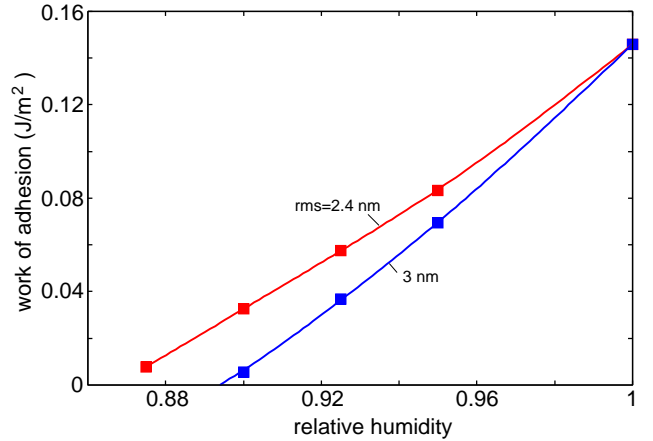}
\caption{\label{theory.RH.2w}
The work of adhesion as a function of the relative humidity. 
In the calculation we have assumed
hard surfaces with the root mean square roughness $h_{\rm rms} = 2.4$ and $3 \ {\rm nm}$
in contact with an elastic solid (with the Young's modulus $E=82 \ {\rm GPa}$ and Poisson ratio
$\nu = 0.22$) with a flat surface. The square data points are the calculated results
and the lines smoothing cubic splines.
}
\end{figure}

\begin{figure}
\includegraphics[width=0.47\textwidth,angle=0]{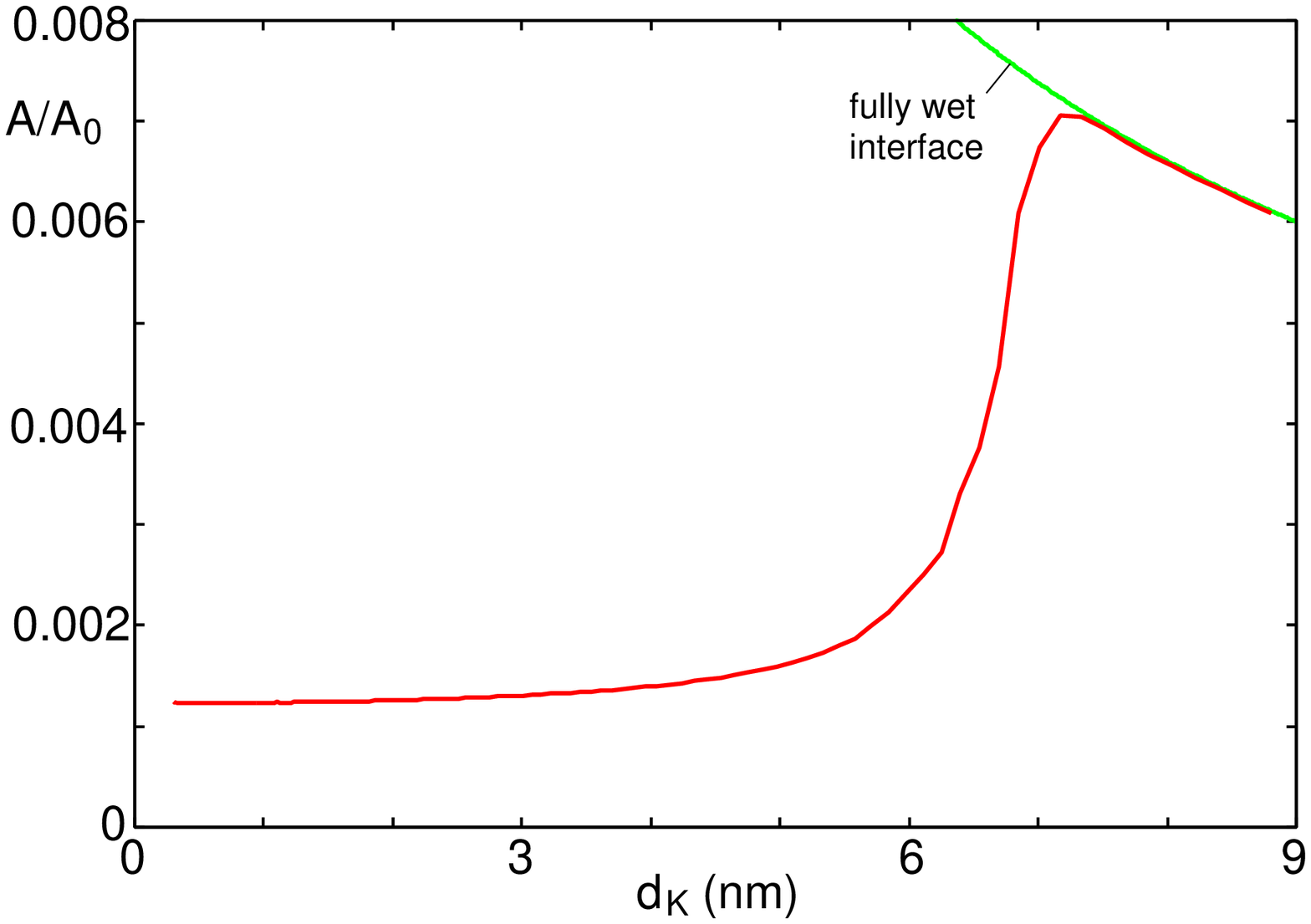}
\caption{\label{DeBoer.area.dK}
The area of real contact
as a function of the Kelvin capillary width $d_{\rm K}$. 
For a hard surface with the root mean square roughness $h_{\rm rms} = 2.4 \ {\rm nm}$
in contact with elastic solids with the Poisson ratio $\nu = 0.22$ and
the Young modulus $E=82 \ {\rm GPa}$.
The (nominal) squeezing pressure $p=4 \ {\rm MPa}$. 
}
\end{figure}

\begin{figure}
\includegraphics[width=0.47\textwidth,angle=0]{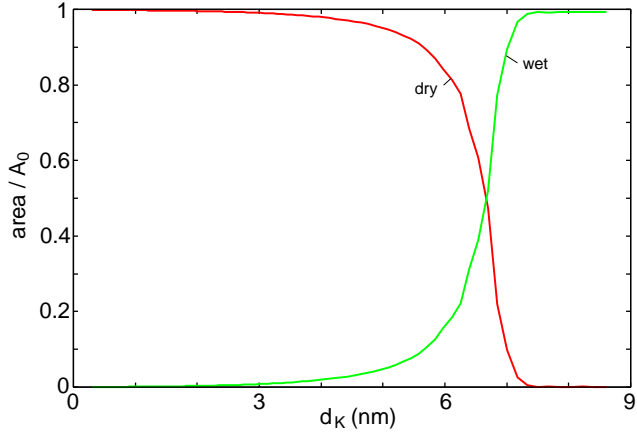}
\caption{\label{De.Boer.area.dry.wet.log.dW}
The dry and wet area 
as a function of the Kelvin capillary width $d_{\rm K}$. 
For a hard surface with the root mean square roughness $h_{\rm rms} = 2.4 \ {\rm nm}$
in contact with elastic solids with the Poisson ratio $\nu = 0.22$ and
the Young modulus $E=82 \ {\rm GPa}$.
The (nominal) squeezing pressure $p=4 \ {\rm MPa}$. 
}
\end{figure}

\vskip 0.3cm
{\bf 3.1 Elastically hard solids}

In Fig. \ref{pull.off.de.Boer}
we show the stress as a function of the (average) distance between the surfaces, $\bar u$, during 
separation for several different relative humidities (RH), 0.95, 0.925, 0.9 and 0.875. 
The elastic solid has the Young's modulus $E=82 \ {\rm GPa}$ and the Poisson ratio
$\nu = 0.22$. 
The area under the curves in Fig. \ref{pull.off.de.Boer}
determines the work of adhesion which is shown in 
Fig. \ref{DeBoer.w.invers.dK}.
The numerical results in Fig. \ref{DeBoer.w.invers.dK} are in relative good agreement with
the (approximate) analytical result (14) (curve denoted by ``approx'' in Fig. \ref{DeBoer.w.invers.dK}).

In Fig. \ref{theory.exp.RH.w} we compare the work of adhesion, as a function of the relative humidity,
with the experimental data of DelRio et al.\cite{Frank1,Frank2}, obtained from 
microcanterlever experiments. 
The circles are experimental data and the squares
calculated results (from Fig. \ref{DeBoer.w.invers.dK}). The experimental system is for polysilicon surfaces 
with the combined surface having the root-mean-square (rms) roughness\cite{Frank2}
$\approx 2.3 \ {\rm nm}$, while in the calculation we have assumed the same system as above 
where the rough surface has the rms roughness $h_{\rm rms} \approx 2.4 \ {\rm nm}$.

In Fig. \ref{theory.RH.2w}
I show the calculated work of adhesion as a function of the relative humidity. 
In the calculation I have assumed 
hard surfaces with the root mean square roughness $h_{\rm rms} = 2.4$ (from Fig. \ref{theory.exp.RH.w})
and $3 \ {\rm nm}$. The power spectrum of the second surface was obtained
from the original power spectrum (bottom curve in Fig. \ref{Cq.both}) by scaling it with a factor
of $(3/2.4)^2$.
The slope of the ${\rm rms}=3\ {\rm nm}$ line is about 25\%
larger than for the ${\rm rms} = 2.4 \ {\rm nm}$ curve. This is just the ratio between the two
rms values and agree with the prediction of the (simplified) theory, Eq. (14),
which shows that the slope of the work of adhesion curve scales approximately
linearly with the rms-value. This prediction is very different from the GW theory prediction
which predict much larger change in the work of adhesion.

In the present case the theory predicts that the 
work of adhesion per unit area, $w$, vanishes for ${\rm RH} < 0.87$ (see Fig. \ref{theory.exp.RH.w}) 
or, equivalently, for the
capillary heights $d_{\rm K} < 8 \ {\rm nm}$ (see Fig. \ref{DeBoer.w.invers.dK}). In the experiment $w$ 
is indeed very small for ${\rm RH} < 0.87$ but not zero (see Fig. \ref{theory.exp.RH.w}), 
and we attribute this difference
between theory and experiment to two different effects:

1) Finite size effect: The theory is for an infinite system 
while the experimental system is finite, and in the present case in fact quite small (the crack tip
process zone has a total area of only $\sim 100 \ {\rm \mu m}^2$, see Ref. \cite{Boer}). A finite pull-off force
due to capillary bridges will in reality 
always occur, even in the limit of very low RH where (for a small system and solids
which high elastic modulus) 
a single asperity (or just a few asperities) may be in contact--in this case
a (small) capillary bridge can form at the asperity 
giving a non-zero pull-off force. For a finite-sized system this will
result in a non-zero work per unit area, $w$, to separate the solids 
(which may be non-negligible for micro and nano-sized systems), but
in the thermodynamic limit (infinite sized system, as assumed in the theory), $w$ would of course vanish.
The small contact between small, elastically hard, solids also imply 
that there will be large fluctuations in the
pull-off force between different realizations of the same system. 
This has indeed been observed by Zwol et al\cite{Zwol}, who
found that the pull-off force (at low relative humidity) 
varied by a factor of $\sim 2$ (or more), 
when the same microsized object was brought into contact
with the same substrate surface at different locations (see also Sec. 5).

2) Van der Waals interaction: 
One can easily show that the long-ranged van der Waals interaction will
always gives a non-zero work of adhesion. 
The reason for this is that the (repulsive) 
contribution to the wall-wall interaction
from the elastic deformation of asperities decay as $\sim {\rm exp}(-\alpha \bar u/h_{\rm rms})$ 
(or faster, if finite size effects are taken into account) with
increasing wall separation $\bar u$, while the (attractive) van der Waals interaction
decay slower as $\sim 1/\bar u^3$
and the total interaction will always be attractive for large enough separation $\bar u$.  
With only the capillary bridges
the wall-wall interaction is predicted to be repulsive (for all wall-wall separations)
at low relative humidity. 
Combining the van der Waals interaction and the capillary
contribution will give a non-zero work of adhesion even for small 
relative humidity, which still depend on the relative humidity.
I plan to study this in more detail in the future.  

It is interesting to compare (14) with the predictions of the Greenwood--Williamson (GW) theory of contact
mechanics. As pointed out before, this theory fails qualitatively when 
roughness occurs on many length scales, and  
DelRio et al. found that
the GW theory strongly underestimate the work of adhesion.
The main reason for this failure is not the asperity approximation (which, however, also is a severe
approximation in the present case) but the neglect of elastic coupling between the contact regions.

The separation of two solids usually occurs via interfacial crack propagation, 
which depends on the work of adhesion. 
However, the sliding friction is determined mainly by the area of real contact.
In Fig. \ref{DeBoer.area.dK} we show the area of real contact
as a function of the Kelvin length $d_{\rm K}$.
The (nominal) squeezing pressure is $p=4 \ {\rm MPa}$.  
Note that the area of real contact is maximal at $d_{\rm K} \approx 7 \ {\rm nm}$, 
which correspond to the point where there is just enough fluid to fill
the space between the solids. 
This is illustrated in Fig. \ref{De.Boer.area.dry.wet.log.dW} which shows the wet (and dry) area 
as a function of the Kelvin length $d_{\rm K}$. Note that for 
$d_{\rm K} > 7 \ {\rm nm}$ the interface if filled with liquid. In this case the
negative pressure $2 \gamma / d_{\rm K}$ will prevail (nearly) everywhere at the interface
and the area of real contact will therefore be proportional to $1/d_{\rm K}$, which is
indeed the dependence of $A/A_0$ on $d_{\rm K}$ observed in Fig. \ref{DeBoer.area.dK} for
$d_{\rm K} > 7 \ {\rm nm}$. 
In fact, we can accurately describe the behavior of the contact area for large and small
$d_{\rm K}$. Thus, when $d_{\rm K} = 0$ no liquid occur at the interface, and since the 
applied (squeezing) pressure $p_0$ is very small the area of real contact is linearly related to the $p_0$
according to 
$${A \over A_0} = \alpha p_0$$
where $\alpha$ can be calculated from the surface roughness power spectra and the effective
elastic modulus as described elsewhere\cite{JCPpers,PSSR}. 
We get $\alpha = 3.1 \times 10^{-10} \ {\rm Pa}^{-1}$
which gives the limit $A/A_0 \approx 1.24\times 10^{-3}$ as $d_{\rm K} \rightarrow 0$, in good agreement
with Fig. \ref{DeBoer.area.dK}.

In the opposite limit of large $d_{\rm K}$ (nearly) the whole interface is filled by fluid so that the
effective pressure acting on the block is the sum of the applied pressure
$p_0$ and the capillary pressure $2\gamma / d_{\rm K}$ giving
$${A\over A_0} = \alpha \left ({2 \gamma \over d_{\rm K}}+p_0\right )$$
This relation (denoted ``fully wet interface'')
is shown in Fig. \ref{DeBoer.area.dK} and agree very well with the numerical result
for $d_{\rm K} > 7 \ {\rm nm}$.

\begin{figure}
\includegraphics[width=0.47\textwidth,angle=0]{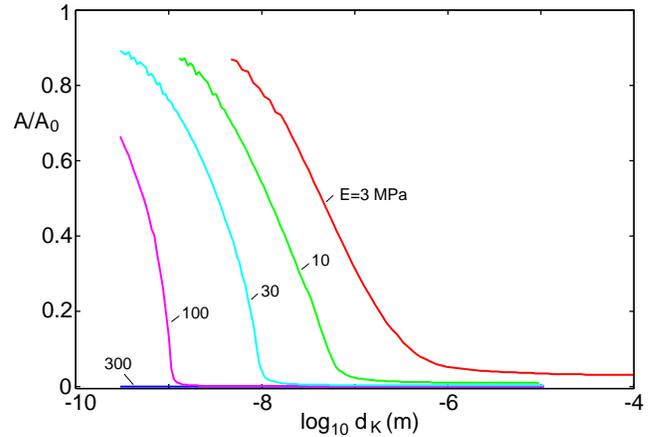}
\caption{\label{Rubber.log.dK.area.contact}
The area of real contact as a function of the logarithm of the Kelvin capillary width $d_{\rm K}$ 
For a hard surface with the root mean square roughness $h_{\rm rms} = 6 \ {\rm \mu m}$
in contact with elastic solids (with a flat surfaces) with the Poisson ratio 0.5 and several different
Young modulus, $E=3$, $10$, $30$, $100$ and $300 \ {\rm MPa}$. 
The (nominal) squeezing pressure $p=0.1 \ {\rm MPa}$. 
}
\end{figure}

\begin{figure}
\includegraphics[width=0.47\textwidth,angle=0]{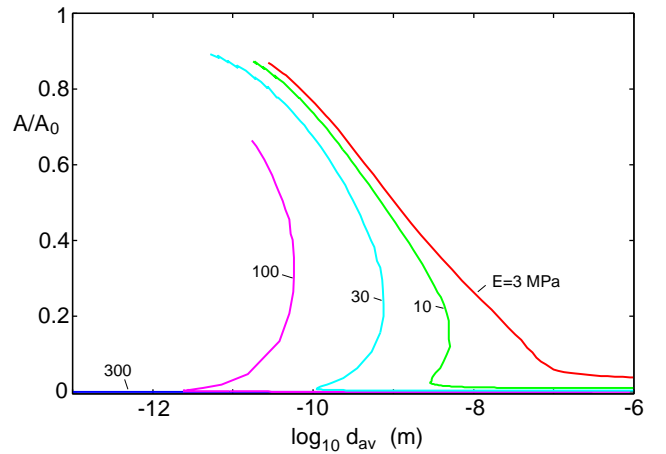}
\caption{\label{Rubber.log.daverage.contact.area}
The area of real contact as a function of the logarithm of the average
width, $d_{\rm av}$, of the water layer between the surfaces.
For a hard surface with the root mean square roughness $h_{\rm rms} = 6 \ {\rm \mu m}$
in contact with elastic solids with the Poisson ratio 0.5 and several different
Young modulus, $E=3$, $10$, $30$, $100$ and $300 \ {\rm MPa}$. 
The (nominal) squeezing pressure $p=0.1 \ {\rm MPa}$.
}
\end{figure}


\begin{figure}
\includegraphics[width=0.47\textwidth,angle=0]{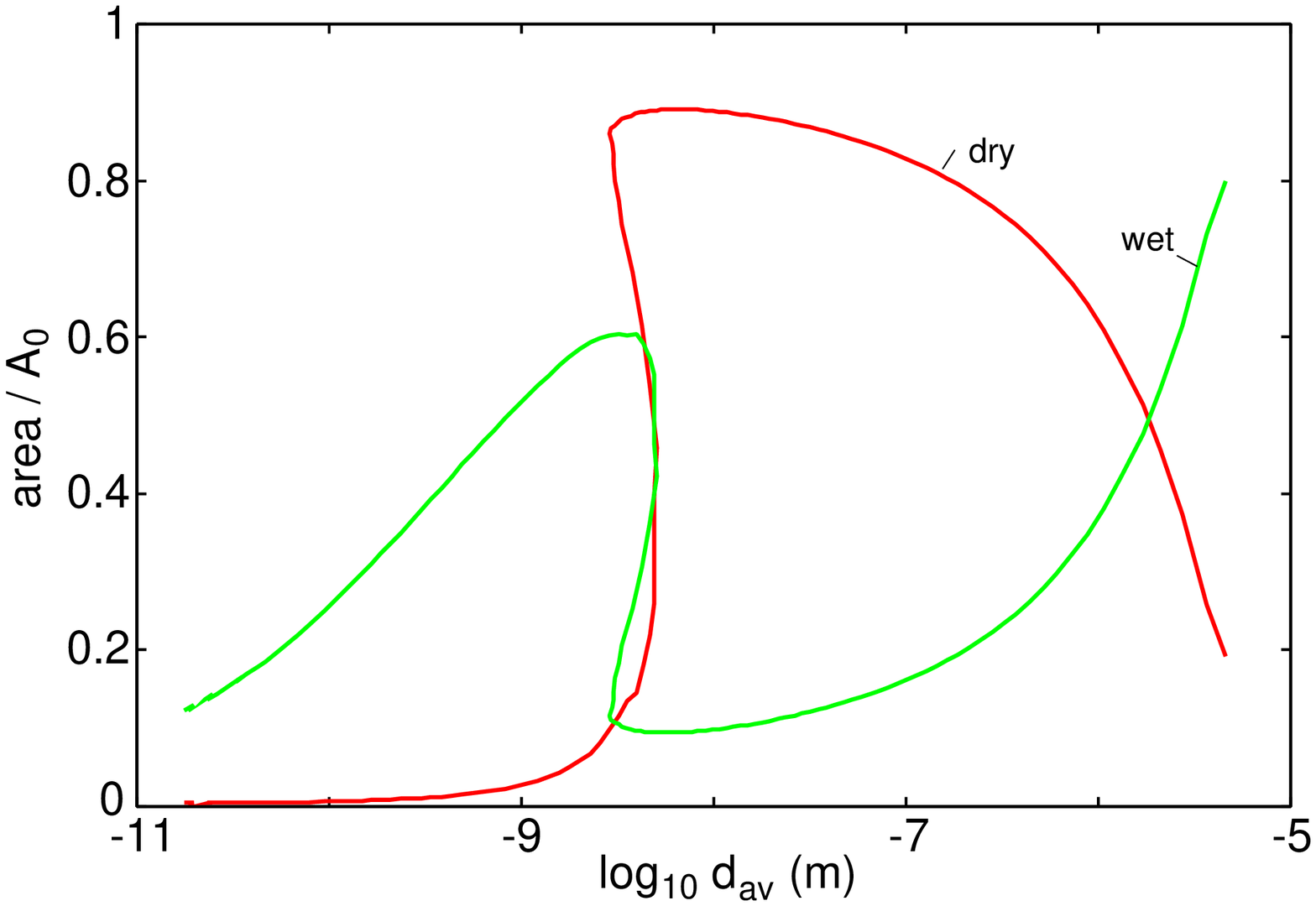}
\caption{\label{Rubber.E=10MPa.wet.dry.contact2}
The dry and wet area 
as a function of the logarithm of the average
width, $d_{\rm av}$,  of the water layer between the surfaces. 
For a hard surface with the root mean square roughness $h_{\rm rms} = 6 \ {\rm \mu m}$
in contact with elastic solids with the Poisson ratio 0.5 and
the Young modulus $E=10 \ {\rm MPa}$.
The (nominal) squeezing pressure $p=0.1 \ {\rm MPa}$. 
}
\end{figure}


\begin{figure}
\includegraphics[width=0.47\textwidth,angle=0]{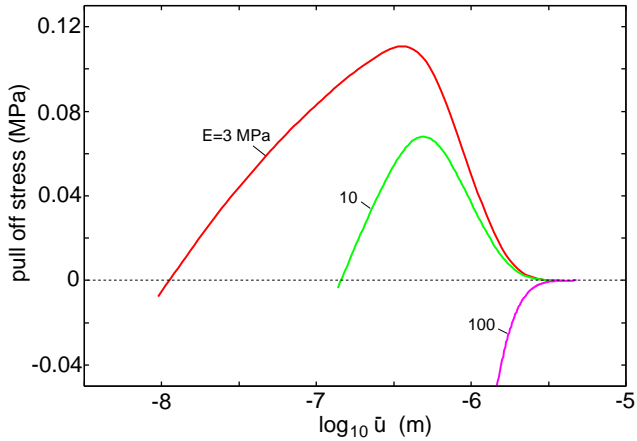}
\caption{\label{Rubber.log.baru.pull.off.stress}
The stress as a function of the (average) separation $\bar u$ during 
separation. The area under the curves (for positive pull-off stress)
determines the work of adhesion. For
a hard surface with the root mean square roughness $h_{\rm rms} = 6 \ {\rm \mu m}$
in contact with an elastic solids (with the Young's modulus $E=3$, $10$ and $100 \ {\rm MPa}$ 
and Poisson ratio
$\nu = 0.5$) with a flat surfaces. Results are shown for the Kelvin 
distance $d_{\rm K} = 1 \ {\rm \mu m}$
(corresponding to the relative humidity ${\rm RH} \approx 0.999$). 
}
\end{figure}

\vskip 0.3cm
{\bf 3.2 Elastically soft solids}

Consider the contact between an elastically soft solid with a flat surface and a hard, 
randomly rough, substrate with the 
power spectrum shown in Fig. \ref{Cq.both}, upper curve.   
This system exhibit more complex and interesting behavior than the case of elastically hard
solids studied in the last section. Thus, we find that when the relative humidity decreases,
resulting in a higher negative pressure in the capillary bridges, there may be a strong
increase in the area of real contact because of elastic deformations of the solid walls,
which are pulled into closer contact by the capillary adhesive forces. This is in sharp contrast to the
case of hard solids (with relative smooth surfaces) studied above, 
where the contact area was maximal when there was just enough liquid 
to fill the space between the (almost) undeformed solid walls.

Fig. \ref{Rubber.log.dK.area.contact} shows 
the area of real contact as a function of the logarithm of the Kelvin capillary width $d_{\rm K}$, 
for a hard surface with the root mean square roughness $h_{\rm rms} = 6 \ {\rm \mu m}$,
in contact with elastic solids (with flat surfaces) with the Poisson ratio 0.5, and several different
Young modulus, $E=3$, $10$, $30$, $100$ and $300 \ {\rm MPa}$.
The (nominal) squeezing pressure $p=0.1 \ {\rm MPa}$. 
Note that for low enough RH (i.e., for small enough capillary distance $d_{\rm K}$)  the area
of real contact increases rapidly with decreasing RH. This result from the strong capillary bridges
pulling the solids into close contact. As expected, this effect start at higher and higher RH as the elastic
modulus of the solids decreases. 
For the stiffest solid, $E=300 \ {\rm MPa}$, the area of real contact does not increase
when the RH is reduced even to the point that the height of the fluid bridges is of molecular size 
$d_{\rm K} \approx 0.5 \ {\rm nm}$ (below which the continuum theory fail).

Fig. \ref{Rubber.log.daverage.contact.area}
shows again the area of real contact but now as a function of the logarithm of the average
width, $d_{\rm av}$, of the water layer between the surfaces. This figure illustrates an interesting effect:
as the RH decreases, in some range of RH the amount of 
liquid between the surfaces increases as the RH decreases, i.e., water
from the atmosphere condenses at the interfacial region between the solids. The reason
for this is that when the RH decreases, the surfaces are pulled together which imply that the wet interfacial
area [where $u({\bf x}) < d_{\rm K}$] 
increases, and this increase is so fast that the volume of liquid between the surfaces increases in spite
of the fact that the capillary height decreases.

This effect is also illustrated in 
Fig. \ref{Rubber.E=10MPa.wet.dry.contact2}
which show the wet (and dry) area 
as a function of the logarithm of the average
width, $d_{\rm av}$,  of the water layer between the surfaces. 

In Fig. \ref{Rubber.log.baru.pull.off.stress}
we show the stress as a function of the (average) interfacial separation $\bar u$, during 
pull-off. The area under the curves (for positive pull-off stress) determines the work of adhesion. 
Results are shown for elastic solids with a flat surfaces, 
and with the Young's modulus $E=3$, $10$ and $100 \ {\rm MPa}$ 
and the Poisson ratio
$\nu = 0.5$. The results are for the Kelvin 
distance $d_{\rm K} = 1 \ {\rm \mu m}$
[corresponding to the relative humidity (RH) $\approx 0.999$]. 
Note that for the most stiff solid $E=100 \ {\rm MPa}$
the interaction is purely repulsive, i.e., the work of adhesion vanish.
For the elastically softer solids the work of adhesion is finite, but the pull-off stresses 
are much smaller than for the hard smooth surfaces studied
in Sec. {\bf 3.1.}, while the distance over which the pull-off force is non-negligible are much longer, both
effects reflecting the much larger 
fluid film thickness (or Kelvin distance $d_{\rm K}$) in the present case.
The work of adhesion (i.e., the area under the curves in Fig. \ref{Rubber.log.baru.pull.off.stress})
for the $E=3 \ {\rm MPa}$ case is $w=0.11 \ {\rm J/m^3}$ and for the 
$E=10 \ {\rm MPa}$ case we get $w=0.062 \ {\rm J/m^3}$.

\vskip 0.5cm
{\bf 4. Applications}

We present a few applications of the results presented above. Consider first
rubber friction on a wet glass surface in the context of wiper blades. After rain or
car wash, the driver of a car does not immediately stop the motion of the wiper blades.
In this case, as the water is removed by wiping and evaporation, as a function of time a high friction peak,
with a friction considerably higher than the dry one, may be observed.
This time period of enhanced friction is denoted as the {\it tacky regime}, 
and may prevail for several seconds. 
Experiment\cite{tacky1} have shown that in the tacky regime the attraction 
from water capillary bridges between the rubber and the glass
substrate pull the rubber into close contact with the substrate 
so that the area of real contact is even larger in the tacky
regime than for the perfectly dry contact. Since the 
applied pressure in the rubber-glass nominal contact area is very high in
wiper blade applications, typically of order MPa, 
the additional contribution from the capillary bridges must be very large, 
of order MPa, in order to explain the 
strong increase (typically by a factor of 2) in the friction. This implies that the height of
the capillary bridges $d < 100 \ {\rm nm}$. 
This also implies that the wiper blade-glass rubber friction may be considerably enhanced
at humid condition, where capillary bridges are formed spontaneously. 

The hight friction in the tacky region can block a wiper system. Capillary adhesion can be reduced
by surface treatment of the rubber. Thus, halogenation result in a rubber surface layer which is elastically
much stiffer than the bulk (by up to a factor of 100, see Ref. \cite{Koenen1})
and will strongly reduce the friction coefficient
in the tacky region, e.g., from 3
(no halogenation) to 1 (at $15 \%$ halogeneous concentration in the surface region)\cite{Koenen}.
The surfaces of wiper blades are almost always exposed to halogenation. However, after long time of use
this layer of modified rubber may be removed by wear. Thus, the friction coefficient for very worn-out 
blades may exceed 3.

Capillary bridges may also give 
rise to strongly enhanced friction for lubricated rubber applications, if the lubrication film is very
thin everywhere (e.g., even at the boundary of the nominal contact area). 
Thus, in a recent experiment\cite{Kroger}, 
a rubber block was slid on a smooth steel surface lubricated by a drop of oil.
In this case the oil film where everywhere very thin $d \approx 300 \ {\rm nm}$, 
and the rubber-steel surface friction was observed to increase from
$\sim 1$ to $\sim 3$ when the oil drop was added to the metal surface. 
In this experiment the applied nominal pressure was rather small,
about $0.01 \ {\rm MPa}$.
The capillary adhesive force can be estimated
using $2 \gamma /d \approx 0.3 \ {\rm MPa}$ where we used $\gamma \approx 0.05 \ {\rm J/m^2}$. 
This is much larger than the applied nominal squeezing pressure, but because of the finite sliding speed
($0.1 \ {\rm m/s}$) one cannot assume that all the fluid get completely 
squeezed out from the asperity contact regions\cite{Mugele}, and
therefore the increase in the sliding friction will be smaller than indicated by the increase in the 
effective normal load. 

Capillary adhesion is very important in biological adhesive 
systems used for locomotion. This has been studied in
details for tree frogs\cite{frog1,frog2} and stick 
insects\cite{Federle1}. These animals use smooth adhesive pads which are built from non-compact materials
and are elastically very soft. To adhere to surfaces the animals 
inject a wetting liquid into the contact area. Experiments for
stick insects have shown that the nominal 
frictional shear stress for sliding against smooth glass surfaces increases when the fluid
film thickness decreases, e.g., during sliding long distances. During repeated sliding on the same surface area, the friction decreased
continuously as more liquid accumulated on the 
surface -- typically the frictional shear stress dropped from $0.15 \ {\rm MPa}$ to
$0.05 \ {\rm MPa}$ as the film thickness increased. 
Nevertheless, as expected, the nominal pad-substrate surface contact area did not change much
since the work of adhesion (which determines the contribution to the 
contact area from the adhesional interaction)
may be nearly constant (equal to $2\gamma$) if enough liquid occurs at the interface, as
indeed expected in the present case (note:
the insects usually need to move on surfaces much rougher than the glass surface and
must therefore inject much more liquid at the pad-substrate 
contact area than is necessary for the smooth glass surface).
The pull-off force from the glass surface was nevertheless observed 
to increase when the film thickness decreased during repeated 
contact and pull-off. This is most likely a viscosity effect: 
during separation of two closely spaced solids separation by a thin
fluid layer, fluid must flow towards the center of the contact area, 
and during ``fast'' pull-off this generates a strong negative pressure
(effective adhesion) in the fluid because of viscous
dissipation in the fluid. In this case, the thinner the fluid film, the larger peak force will be, 
as seen in the experiments.

As a final application of the theory presented above, let us consider the contact
between the toe pad of a tree frog and a smooth glass plate. This system has been studied in great
detail in Ref. \cite{frog1}. At the toe pad-substrate interface occurs a wetting liquid, which is likely
to be of order $\sim 1 \ {\rm \mu m}$ thick 
at the edges of the toe pad--substrate contact regions (see Ref. \cite{frog2}).
Thus, the negative capillary pressure $2\gamma / d \approx 0.1 \ {\rm MPa}$. 
Because of the low effective elastic modulus of the toe pad, 
it deforms elastically so that a large fraction (about $50 \%$) of the toe pad
comes into close contact (less than $5 \ {\rm nm}$) with the glass surface, in spite of the fact that the
natural height fluctuations of the toe pad surface (on the length scale of $\sim 10 \ {\rm \mu m}$) 
is of order $\sim 1 \ {\rm \mu m}$.
The area of real contact manifests itself experimentally as a finite static friction force.

\vskip 0.5cm
{\bf 5. Discussion}

The theory presented in Sec. 2 is a mean-field type of theory, 
where the adhesive force $F_{\rm a}$ from the capillary bridges is taken
into account by adding an adhesive pressure 
$p_{\rm a}=F_{\rm a}/A_0$ to the external squeezing pressure $p_0 = F_0/A_0$ acting on the 
block ($F_0$ is the normal load, which is assumed to act 
uniformly on the top surface of the block). This is likely to be an accurate
approximation as long as a large fraction of the non-contact 
interfacial region is filled with liquid, e.g., for high relative 
humidity. However, for very low relative humidity and for large 
surface roughness, liquid bridges may only occur close to the
outer edges of the asperity contact regions. If the regions 
occupied by fluid are very small compared to the diameter of the
contact and non-contact regions, then we can consider the fluid filled regions as crack-tip process zones, 
and this limit can be studied using the adhesion theory developed elsewhere\cite{P1,Pere}. In this theory 
the interface is studied at different magnifications, and an effective interfacial
binding energy (per unit area), $\gamma_{\rm eff}(\zeta)$, is introduced. If one assumes that
the continuum theory of capillary forces 
is still valid for the small capillary bridges which occur at low RH then
$\gamma_{\rm eff}(\zeta_1) = 2\gamma {\rm cos}\theta$. However, most likely the continuum theory 
is not valid for narrow (molecular sized) capillary
bridges, and one may have to
treat $\gamma_{\rm eff} (\zeta_1) = \Delta \gamma$ as an 
empirical parameter to be determined directly from experimental data.   

In the theory developed in Sec. 2 we always assumed thermal 
equilibrium. However, it is known that the formation of capillary
bridges in a humid atmosphere is a thermally activated process, 
with a continuum of activation energies and hence relaxation times.
Thus, in general it may take very long time to reach (or come close to) thermal equilibrium. In the study by 
Maarten et al\cite{Boer} for
microcantilevers, at the relative humidity 0.3, the average crack length 
(defined as the non-contact part of the cantilever) decreased with increasing times
from $\approx 700 \ {\rm \mu m}$ 
initially to $\approx 550 \ {\rm \mu m}$ after $2.5 \ {\rm h}$, to $\approx 400 \ {\rm \mu m}$
after $5 \ {\rm h}$, and to $\approx 375 \ {\rm \mu m}$ 
after $10 \ {\rm h}$. After this time no further decrease of the crack
length was observed. In general, the increase in 
adhesion and friction with increasing time, as a result of thermally activated
formation of capillary bridges, 
has been studied in detail, and is of great importance in many systems of fundamental or applied 
interest\cite{Boc,Riedo,BookP}.   

Another limitation of the theory presented in Sec. 2 
is that it is only valid if most of the relevant repulsive
wall-wall interaction occurs for separations $\bar u$ 
that are small enough so that, for the actual physical system,
there are enough asperity contact regions 
(to obtain enough self-averaging) that the analytical theory can be applied. 
That is, the analytical contact mechanics theory is for an infinitely
system (thermodynamic limit), and for a randomly rough surface 
(with a Gaussian probability height distribution)
there will always be infinite high asperities, and contact 
between two solids will occur for any (average) separation
$\bar u$ between the solids. In fact, for large $\bar u$ we have the exact result
$p \sim {\rm exp}(-\alpha \bar u/h_{\rm rms})$, 
which shows that a repulsive pressure $p$ will act between the
solids for arbitrary large  separation $\bar u$. However, for  
finite-sized systems, the highest asperities have a finite
height, which for macroscopic systems typically is of order 
$\sim 10 h_{\rm rms}$ (see Appendix A in Ref. \cite{PSSR}), 
but in the context of MEMS may be
much smaller due to their small physical size. 
In fact, de Boer has estimated the highest asperities in the crack tip 
process zone to be only $\approx 3.7 h_{\rm rms}$. 
It is clear that the analytical theory is only valid if the important
repulsive contribution to the work of adhesion occurs for separations $\bar u < 3.7 h_{\rm rms}$. 
In our applications to microcantilever adhesion most of the relevant repulsive interaction 
occurs for $\bar u < 3.2 h_{\rm rms}$ so the theory is likely to be at least semiquantitatively correct.

\begin{figure}
\includegraphics[width=0.47\textwidth,angle=0]{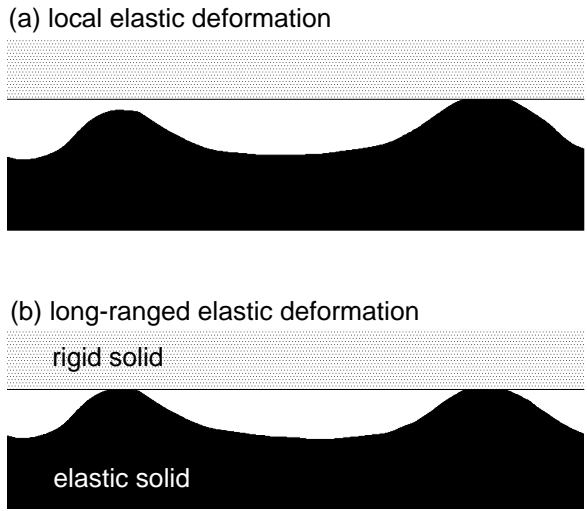}
\caption{\label{local.longranged}
(a) In the Greenwood--Williamson (GW) contact mechanics model
only local (asperity) elastic deformations are included.
(b) In reality the elastic deformation is long ranged, and asperities 
which never could be in contact
in the GW theory because they are too low, 
could be in contact when the elastic deformations is fully included in the analysis.
For this reason the GW theory is much more sensitive to the exact form of the probability distribution
of asperity heights for large asperity height, than in a more accurate analysis which include 
the long-range elastic deformations.
}
\end{figure}

In the GW theory a similar 
problem related to the height of the highest asperities, but in addition the exact form of the 
tail of the height distribution for large separation 
matters a lot for the contact mechanics and in particular for the relation
between $\bar u$ and $p$. 
However, this is an artifact of the GW theory, and when the long-range elastic coupling is
taken into account as in the theory of Persson, 
the contact mechanics becomes much less dependent on the details of the
height distribution. This has been verified by 
Finite Element Method calculations\cite{Pei} for polymer surfaces\cite{Benz} where, in spite of the
fact that the height probability is rather non-Gaussian for large asperity heights 
(see Fig. 14 in Ref. \cite{YangPersson}), the contact mechanics obeys
the usual behavior with the area of real contact 
proportional to the load and the average interfacial separation depending on the
squeezing pressure as $p \sim {\rm exp}(-\alpha \bar u/h_{\rm rms})$, in good numerical agreement with the 
Persson theory\cite{YangPersson}. This can be 
explained by the fact that because of long range elastic deformation, not only the asperities
close to the top of the Gaussian distribution will make contact, 
but a larger range of asperity heights will be involved (which is the
reason for why the the asymptotic relation $p\sim {\rm exp}(- \alpha \bar u/ h_{\rm rms})$ is exponential rather than Gaussian (reflecting
a Gaussian height distribution)
$p\sim {\rm exp}[-b (\bar u / h_{\rm rms})^2]$ 
as in the GW theory). This is illustrated in Fig. \ref{local.longranged} for
a situation where this fact is particularly clear.

The analytical theory presented in Sec. 2 for the limiting case of an interface 
completely filled with fluid (e.g., high
relative humidity) indicates that the most important property 
of the rough surface is the root-mean-square roughness $h_{\rm rms}$,
while other properties such as (for self affine fractal surfaces) 
the fractal dimension $D_{\rm f}$ and the upper and
lower cut off wavevectors $q_1$ and $q_0$ are much less important because the parameter $\alpha$ in the
relation $p\sim {\rm exp}(-\alpha \bar u /h_{\rm rms})$ 
is very insensitive to these quantities\cite{PerssonPRL}. This is likely to be the
case also for non-fractal surfaces. 
This fact is consistent with experiments of van Zwol et al\cite{Zwol}, who observed
that surfaces with very different $q_0$ but the same root-mean-square roughness 
$h_{\rm rms}\approx 1.5 \ {\rm nm}$, exhibited the same capillary adhesion. 
We conclude that the rms roughness gives the
predominant effect on the adhesive force due to capillary bridges.

In the theory presented in Sec. 2 we have neglected plastic deformation of the 
solids and the disjoining pressure due to
adsorbed (water) layers. The problem of the influence of plastic yielding on the contact mechanics was
studied in Ref. \cite{Frank2} where it was found to have a 
very small influence on the capillary adhesion. I have analyzed the same
problem using the more accurate contact 
mechanics theory I have developed. I find that the surface roughness with wavelength 
longer than $\sim 0.2 \ {\rm \mu m}$ undergoes negligible plastic deformation, while some plastic deformation
occurs for roughness at shorter length scales. Plastic deformation will tend to smoothen the surfaces and
hence reduce the stored elastic energy. This in turn will enhance the capillary adhesion. I will report
on this study elsewhere.
 
The presence of adsorbed water layers on the hydrophilic polysilicon 
surfaces used in the experiment\cite{Frank2} will also
play a role in the capillary adhesion process. Unfortunately, 
a water adsorption isotherm does not currently exist for polysilicon surfaces,
which makes it impossible to definitively determine the role of 
adsorbed water layers on the capillary adhesion problem. However,
the fact that we obtained good agreement with experiment neglecting 
this effect for high relative humidity, indicates that it may not be very important 
(for ${\rm RH} > 0.9$) for the
microcantilever applications discussed above.

\vskip 0.5cm
{\bf 6. Summary and conclusion}

I present a general theory for how the contact area and the work of adhesion 
between two elastic solids with randomly rough surfaces
depends on the relative humidity. The surfaces are assumed to be hydrophilic, and
capillary bridges form at the interface between the solids. For elastically hard solids with
relative smooth surfaces, the area of real contact and therefore also the sliding friction,
are maximal when there is just enough liquid to fill out the 
interfacial space between the solids, which typically occurs for $h_{\rm K} \approx 3 h_{\rm rms}$,
where $h_{\rm K}$ is the height of the capillary bridge and $h_{\rm rms}$ the root-mean-square roughness of
the (combined) surface roughness profile.
For elastically soft solids, the area of real contact is maximal for very low humidity where
the capillary bridges are able to pull the solids into nearly complete contact. In both case,
the work of adhesion is maximal (and equal to $2\gamma {\rm cos}\theta$, where $\gamma$ is the liquid
surface tension and $\theta$ the liquid-solid contact angle) 
when $d_{\rm K} >> h_{\rm rms}$, corresponding to high relative humidity.

The theory is compared to experimental data for microcantilever structures. The theory is in good agreement with the
data while the classical Greenwood-Williamson theory fail qualitatively. I also present applications
to rubber wiper blades, where the theory can explain the large friction observed in the 
so called ``tacky region'' for nearly dry contacts, 
where the capillary bridges pulls the rubber into intimate contact
with the glass substrate over a large fraction of the nominal contact area.    

\end{document}